\documentclass[12pt]{iopart}

\usepackage{graphicx}
\usepackage{color}
\usepackage{psfrag}
\usepackage[small]{caption}

\definecolor{lgray}{gray}{.80}

\newcommand{\text}[1]{\ensuremath{\mathrm{#1}}}

\newcommand{\ket}[1]{\,\big|{#1}\big> }

\newcommand{\matrixe}[3]{\big< \,{#1}\, \big| \,{#2}\, \big| \,{#3}\, \big> }

\newcommand{\op}[1]{\hat{\mathrm{#1}}}

\newcommand{\conop}[1]{\op{#1}^{\dagger}}
\newcommand{\desop}[1]{\op{#1}^{\phantom{\dagger}}}

\newcommand{\linemediumsolid}[1][black]{\unitlength1ex
	  ({\color{#1}\begin{picture}(6,1)
	     \linethickness{0.4mm}
	       \put(0,0.5){\line(1,0){6.0}}
	         \end{picture}})\nolinebreak
}

\newcommand{\linemediumdashed}[1][black]{\unitlength1ex
	  ({\color{#1}\begin{picture}(6,1)
	     \linethickness{0.4mm}
	       \put(0,0.5){\line(1,0){1.5}}
	         \put(2.2,0.5){\line(1,0){1.5}}
		   \put(4.4,0.5){\line(1,0){1.5}}
		     \end{picture}})\nolinebreak
}

\newcommand{\linemediumdotted}[1][black]{\unitlength1ex
	  ({\color{#1}\begin{picture}(6,1)
	     \linethickness{0.4mm}
	       \put(0,0.5){\line(1,0){0.4}}
	         \put(0.9,0.5){\line(1,0){0.4}}
		   \put(1.8,0.5){\line(1,0){0.4}}
		     \put(2.7,0.5){\line(1,0){0.4}}
		       \put(3.6,0.5){\line(1,0){0.4}}
		         \put(4.5,0.5){\line(1,0){0.4}}
			   \put(5.4,0.5){\line(1,0){0.4}}
			     \end{picture}})\nolinebreak
}

\newcommand{\linemediumdashdot}[1][black]{\unitlength1ex
	  ({\color{#1}\begin{picture}(6,1)
	     \linethickness{0.4mm}
	       \put(0,0.5){\line(1,0){0.4}}
	         \put(0.9,0.5){\line(1,0){1.5}}
		   \put(2.9,0.5){\line(1,0){0.4}}
		     \put(3.8,0.5){\line(1,0){1.5}}
		       \put(5.8,0.5){\line(1,0){0.4}}
		         \end{picture}})\nolinebreak
}

\begin{document}

\title{Response of Bose gases in time-dependent optical superlattices}

\author{Markus Hild, Felix Schmitt, and Robert Roth}

\address{Institut f\"ur Kernphysik, Technische Universit\"at Darmstadt, 64289 Darmstadt, Germany}

%\ead{markus.hild@physik.tu-darmstadt.de}

\begin{abstract}
The dynamic response of ultracold Bose gases in one-dimensional optical lattices and superlattices is investigated based on exact numerical time evolutions in the framework of the Bose-Hubbard model. The system is excited by a temporal amplitude modulation of the lattice potential, as it was done in recent experiments. For regular lattice potentials, the dynamic signatures of the superfluid to Mott-insulator transition are studied and the position and the fine-structure of the resonances is explained by a linear response analysis. Using direct simulations and the perturbative analysis it is shown that in the presence of a two-colour superlattice the excitation spectrum changes significantly when going from the homogeneous Mott-insulator the quasi Bose-glass phase. A characteristic and experimentally accessible signature for the quasi Bose-glass is the appearance of low-lying resonances and a suppression of the dominant resonance of the Mott-insulator phase.   
\end{abstract}


\submitto{\JPB}

\maketitle

%%%%%%%%%%%%%%%%%%%%%%%%%%%%%%%%%%%%%%%%%%%%%%%%%%%%%%%%%%%%%%%%%%%%%%%%%%%%%%%%%%%%%%%%%%%%%%%%%%
\section{Introduction}

Ultracold bosonic gases in optical lattices offer a supreme laboratory for the study of a wide range of phenomena in strongly correlated quantum systems. Due to the flexibility of the experimental methods available for preparing and controlling the relevant parameters, one can address fundamental questions of importance in condensed matter physics and quantum information processing. 

In addition to the powerful experimental tools for preparation and control, several techniques for probing the quantum state of these many-body systems are available. The simplest experimental observable is the matter-wave interference pattern after realease of the atoms from the lattice, which led to the observation of the superfluid (SF) to Mott-insulator phase transition (MI) \cite{GrMa02a,GrMa02b,StMo04,JaBr98,OoSt01,RoBu03a}. Another tool for characterizing the different quantum phases are dynamical excitations. The excitation via temporal lattice modulations corresponding to Bragg spectroscopy \cite{StMo04,StIn99} provides an excellent probe for the observation of quantum phase transitions. In contrast to static perturbations generated, e.g., by tilting the optical lattice \cite{GrMa02a,BrDu04}, Bragg spectroscopy provides a very precise probe for the response at a well-defined excitation energy \cite{StMo04,KBM05,KoIu06,ClJa06}.

Recent experiments also explore the effects of irregular lattice potentials on the phase diagram. One can employ speckle patterns to create random lattice potentials \cite{LyFa05,FoFa05,ScDr05,ClVa05,FaLy06} or use a superposition of two standing-wave lattices of different wavelengths to generate a two-colour superlattice potential \cite{RoBu03b,RoBu03c}.
Recently, the response of the Bose gases in time-dependent superlattices was investigated experimentally by Fallani \emph{et~al.}~\cite{FaLy06}. The spatial structure of the lattice potential gives rise to additional quantum phases, such as localized \cite{Gebh97} and (quasi) Bose-glass phases \cite{RoBu03c}. The systematic experimental investigation of the resulting phase diagram requires powerful probes that allow to distinguish the different phases. In addition to the matter wave interference pattern, the dynamical behaviour of the system provides important information. 

The aim of this paper is to study the response of zero-temperature Bose gases in one and two-colour superlattices and to identify possible experimental hallmarks of the individual quantum phases. To this end, we simulate the time-evolution of the system in the presence of a time-dependent lattice potential in the framework of the Bose-Hubbard model. In section \ref{sec_theory} we formulate the time-dependent Bose-Hubbard Hamiltonian and discuss the numerical approach used for the time-evolution including a physically motivated basis truncation scheme. In preparation of the investigation of superlattice potentials, we study the dynamic response of Bose gases in regular lattice potentials in section \ref{sec_reg_lattice}. In addition to the full time-dependent simulations, we use a linear perturbation analysis \cite{KoIu06,ClJa06} to characterize and interpret the response of the system.  In section  \ref{sec_superlattice} we extend this discussion to two-colour superlattices and identify possible dynamical signatures of the Mott-insulator to quasi Bose-glass transition.

%%%%%%%%%%%%%%%%%%%%%%%%%%%%%%%%%%%%%%%%%%%%%%%%%%%%%%%%%%%%%%%%%%%%%%%%%%%%%%%%%%%%%%%%%%%%
\section{Bose-Hubbard model and time evolution}\label{sec_theory}

\subsection{Bose-Hubbard model}\label{sec_bhm}

The Bose-Hubbard model has proven to be the appropriate framework for the description of zero-temperature Bose gases in optical lattices \cite{JaBr98}. It describes the full phase diagram ranging from weakly interacting superfluid phases to the regime of strong correlations, e.g. in the Mott-insulator phase. Assuming that the lattice potential is sufficiently deep, the single-particle Hilbert space can be restricted to the lowest energy band. A suitable single-particle basis is given by the localized Wannier functions of the lowest band. A state of $N$ bosons at $I$ lattice sites can be represented by an $I$-tuple of occupation numbers $\{n_1,\,n_2,\,\cdots,\,n_I\}$\, of the individual sites \cite{JaBr98,RoBu03a,RoBu03c}. The Fock states $\ket{\{n_1,\,n_2,\,\cdots,\,n_I\}_{\alpha}}$ for all possible compositions of occupation numbers span a basis of the single-band Hilbert space. 

The Bose-Hubbard Hamiltonian for a single-component Bose gas in one dimension reads \cite{JaBr98}:
\begin{equation}
\op{H}=-J\sum_{i=1}^I\left(\conop{a}_{i+1}\desop{a}_i+\text{h.a.}\right)+\sum_{i=1}^I\epsilon_i\op{n}_i+\frac{U}{2}\sum_{i=1}^I\op{n}_i\left(\op{n}_i-1\right)\text{,}\label{eq_hamop}\
\end{equation}
with creation (annihilation) operators~$\conop{a}_{i}$~($\desop{a}_{i}$) for a boson at site~$i$ and occupation number operators~$\op{n}_i=\conop{a}_{i}\desop{a}_{i}$. The first term in (\ref{eq_hamop}) accounts for the hopping between adjacent sites with the tunnelling matrix element $J$. The second term  introduces a site-dependent single-particle energy $\epsilon_i$ which is used to describe, e.g., an external harmonic trapping potential or a superlattice potential \cite{RoBu03b,RoBu03c}. The third term of the Hamiltonian (\ref{eq_hamop}) accounts for the on-site two-body interaction of the atoms with interaction strength $U$. Throughout this paper we use periodic boundary conditions, i.e., hopping between the first and the last site of the lattice is possible.

The parameters of the Bose-Hubbard Hamiltonian are directly connected to the physical lattice realized by an optical standing wave \cite{JaBr98}. The standing wave generates an array of microscopic potentials -- the lattice sites of the Hubbard model. A large intensity of the standing wave, i.e., a deep lattice potential, suppresses the hopping between adjacent sites and confines the atoms to individual sites. At low or vanishing lattice amplitudes the atoms move freely through the lattice and establish long-range coherence.

The dimension $D$ of the number basis for a bosonic system is given by
\begin{equation}
D=\frac{(N+I-1)!}{N!(I-1)!}\text{,}\nonumber
\end{equation}
which increases rapidly with the number of particles $N$ and lattice sites $I$. For fixed filling factor $N/I=1$ the basis dimension of $D=6435$ results for a system of $8$ sites, $D=92378$ for a $10$-site system, and $D=1352078$ for $12$ sites. In the number basis representation the Hamilton matrix is extremely sparse, since only the nearest-neighbour hopping term generates off-diagonal matrix elements. Due to this sparsity the lowest eigenstates can be obtained efficiently  using Lanczos-type algorithms \cite{arpack} for system-sizes up to typically $I=N=12$ with standard desktop computers. The eigenstates are expanded in the number basis,
\begin{equation}
\ket{\nu}=\sum_{\alpha=1}^DC_{\alpha}^{(\nu)}\ket{\{n_1,n_2,\cdots,n_I\}_{\alpha}}\label{,}\label{eq_eigenstate}
\end{equation}
where $C_{\alpha}^{(\nu)}$ denotes the expansion coefficients of the $\nu$-th eigenstate. These coefficients are obtained by diagonalizing the Hamilton matrix.

In order to treat systems of larger size, we have introduced a truncation scheme for the number basis \cite{ScHi06}. In strongly correlated regimes only a few number states contribute to the low-lying eigenstates, since configurations with several atoms occupying a single site are suppressed due to the strong interaction. For the description of the ground state, the basis dimension can be reduced to less than a percent without a significant loss of quality. In contrast, the proper description of a system of weakly interacting atoms requires more number states in the basis. The Hamiltonian itself provides a simple \textit{a priori} measure for the importance of number states via its diagonal matrix elements. Only those states with diagonal matrix elements smaller than a certain truncation energy $E_{\text{trunc}}$, which depends on the Hubbard parameters, are included in the basis:
\begin{equation}
E_{\text{trunc}}\ge\matrixe{n_1,\cdots\ ,n_I}{\op{H}}{n_1,\cdots\ ,n_I}\text{.}
\end{equation}
By adjusting the truncation energy, one can tune the basis for a more accurate description of the physical system or a smaller basis dimension. Detailed benchmarks and applications of this basis truncation scheme can be found in \cite{ScHi06}.

%%%%%%%%%%%%%%%%%%%%%%%%%%%%%%%%%%%%%%%%%%%%%%%%%%%%%%%%%%%%%%%%%%%%%%%%%%%%%%%%%%%%%%%%%%%%%%%%%
\subsection{Time evolution}

Different experimental schemes have been developed to probe the dynamic response of atomic gases in optical lattices. One possible scheme involves the tilting of the lattice to induce a static force \cite{GrMa02a,KBM05}. Another method is Bragg spectroscopy \cite{StIn99}, which has the advantage of not involving any side effects like Bloch oscillations or Zener tunnelling \cite{StMo04,KBM05}. It therefore allows a very precise determination of the excitation spectrum. 

In the case of atoms in an optical lattice this method can be implemented by a temporal modulation of the lattice potential $V_{\text{lat}}(x)$ with frequency $\omega$. A regular one-dimensional optical lattice generated by an optical standing wave is given by 
\begin{equation}
V_{\text{lat}}(x)=V_0\sin^2(kx)\text{,}\label{eq_static_pot}
\end{equation}
with amplitude $V_0$ and wavenumber $k$. The modulation of the lattice is achieved by introducing a time-dependent factor:
\begin{equation}
V_{\text{lat}}(x,t) = V_{0}\left[1+F\sin(\omega t)\right]\sin^2(kx)\label{eq_osc_pot}\text{.}
\end{equation}
The amplitude of the potential (\ref{eq_osc_pot}) is oscillating around the background value $V_0$ with frequency $\omega$ and amplitude $V_0F$ with $F\ll1$. This introduces two sidebands with frequencies $\pm\omega$ and defines the corresponding excitation energy $\omega$ \cite{StMo04}. 

As mentioned in the previous section, the physical lattice (\ref{eq_static_pot}) enters the Bose-Hubbard model via the parameters $J$, $U$, and $\epsilon_i$ \cite{JaBr98}. In order to obtain the time-dependent expressions of these parameters, the localized Wannier states are approximated by Gaussians of width $\sigma$ \cite{KBM05}. The optimal value of the width $\sigma(t)$ is determined by an energy minimization using the lattice potential (\ref{eq_osc_pot}) at a given time $t$. Computing the matrix elements of the kinetic energy and the interaction part of the first quantized Hamiltonian within this Gaussian approximation leads to the time-dependent Hubbard parameters:
\begin{equation}
\begin{array}{rcl}
J(t) & = & J_0 \exp[-F\sin(\omega t)] \;, \\
U(t) & = & U_0 [1 + F\sin(\omega t)]^{1/4} \;.
\end{array}\label{eq_hubbard_params}
\end{equation}
In lowest-order approximation, the temporal change of the on-site energies is directly given by the change of the potential $V_{\text{lat}}(x,t)$, i.e.,
\begin{equation}
\epsilon_i(t) = \epsilon_{i,0}[1+F\sin(\omega t)] \text{.}
\end{equation}
By substituting the static parameters of the Hamiltonian (\ref{eq_hamop}) with the dynamical ones (\ref{eq_hubbard_params}) we obtain the time-dependent Bose-Hubbard Hamiltonian. 

For our simulations of the time evolution in modulated optical lattices, we start with an initial state given by the ground state of the Bose-Hubbard Hamiltonian for given parameters $J_0$, $U_0$, and $\epsilon_{i,0}$.
Starting from this initial state, we then evolve in time steps $\Delta t$ while modulating the lattice with a fixed frequency $\omega$ and amplitude $V_0F$. The frequency $\omega$ defines the probe energy and the response of the system is measured by evaluating observables using the time-evolved state $\ket{\psi,t}$.

%%%%%%%%%%%%%%%%%%%%%%%%%%%%%%%%%%%%%%%%%%%%%%%%%%%%%%%%%%%%%%%%%%%%%%%%%%%%%%%%%%%%%%%%%%%%%%%%%%%%%%%%%%
\subsection{Numerical methods}

The time evolution is performed either by the Crank-Nicholson scheme (CN) or by a 5th order predictor-corrector method\footnote{We use the 5th order Adams-Bashforth predictors and the Adams-Moulton correctors.} (PC), depending on the system parameters. Since the Crank-Nicholson scheme is an implicit method which requires the solution of a set of linear equations\footnote{We employ the PARDISO solver \cite{ScGa00, ScGa04, pardiso} which comes with Intel's \textit{Math Kernel Library}.} of the basis dimension at each timestep, it is feasible only for small systems. Nevertheless, due to its unconditional stability and in combination with the basis truncation, it is a valuable tool to simulate atomic gases in the strongly correlated regime. 

The predictor-corrector method is an explicit scheme which allows to treat larger system sizes. The drawback of this method is its numerical instability, especially in cases where small expansion coefficients of the states (\ref{eq_eigenstate}) are involved. Numerical errors in these coefficients accumulate and lead to a collapse of the computation after a few steps. Nevertheless, it is possible to apply this method in the weakly interacting regime, where only a few small coefficients appear. In connection with the basis truncation, which predominantly discards these number states with small coefficients, the stability is improved. It is, however, not feasible to evolve systems in the strongly interacting regime using the PC method. The initial states consist only of a few dominant number states, i.e., most of the coefficients in the number representation are very small. In these cases the truncation cannot be used to improve the situation, because one would have to neglect a large number of states and, consequently, would miss some features of the excitation spectrum.

%%%%%%%%%%%%%%%%%%%%%%%%%%%%%%%%%%%%%%%%%%%%%%%%%%%%%%%%%%%%%%%%%%%%%%%%%%%%%%%%%%%%%%%%%%%%%%%%%%
\section{Bose gases in an optical lattice}\label{sec_reg_lattice}

\subsection{Setup and linear response analysis}

As a first benchmark, we study the dynamical properties of a Bose gas in a one-dimensional optical lattice with periodic boundary conditions. The system consists of $N=10$ bosons at $I=10$ sites of a regular lattice ($\epsilon_i=0$). The Bose-Hubbard Hamiltonian of such a system is given by
\begin{equation}
\op{H}(t)=-J\sum_{i=1}^I\left(\conop{a}_{i+1}\desop{a}_i+\text{h.a.}\right)+\frac{U}{2}\sum_{i=1}^I\op{n}_i\left(\op{n}_i-1\right)\text{.}\label{eq_reg_hamop}\
\end{equation}

In addition to the exact time evolution, we examine the excitation of the Bose gas using a linear approximation of this Hamilton operator as introduced in refs. \cite{KBM05,IuCa06,KoIu06,ClJa06}. The Hamiltonian (\ref{eq_reg_hamop}) can be written in terms of the hopping and interaction operators, $\op{H}_J$ and $\op{H}_U$, respectively,
\begin{equation}
\op{H}=-J(\tilde{V}_0)\op{H}_J+U(\tilde{V}_0)\op{H}_U\label{eq_short_hamop}
\end{equation}
with the time-dependent amplitude of the physical lattice potential (\ref{eq_osc_pot})
\begin{equation}
\tilde{V}_0(t)=V_0 [1+F\sin(\omega t)].
\end{equation}
We now linearize the Hamiltonian (\ref{eq_short_hamop}) with respect to the perturbation by retaining only the lowest-order terms of a Taylor expansion in the modulation amplitude $F$,
\begin{equation}
\op{H}_{\text{lin}} = \op{H}_0+F\frac{\partial\op{H}}{\partial F}\Bigg|_{F=0}.
\end{equation}
This Hamiltonian depends linearly on the temporal variation of the lattice amplitude. The initial Hamilton operator $\op{H}_0$ is given by (\ref{eq_reg_hamop}) at the time $t=0$.  Evaluation of the derivative of the Hamiltonian~(\ref{eq_short_hamop}) using the dependence of the Hubbard parameters on the lattice amplitude results in

\begin{equation}
\hspace{-10mm}\op{H}_{\text{lin}}(t) = \op{H}_0+FV_0\sin(\omega t)\left[\frac{d\ln U}{d\tilde{V}_0}\Bigg|_{F=0}\op{H}_0-J\left(\frac{d\ln J}{d\tilde{V}_0}\Bigg|_{F=0}-\frac{d\ln U}{d\tilde{V}_0}\Bigg|_{F=0}\right)\op{H}_J\right]\text{,}\label{eq_lin_ham}
\end{equation}
which consists of the unperturbed Hamiltonian $\op{H}_0$ and a linear perturbation part. The first term of the perturbation is  proportional to the unperturbed Hamiltonian and the second term includes the hopping operator $\op{H}_J$. Due to the small modulation amplitude the first part of the perturbation leads to a small energy shift which can be neglected. The second part generates excitations, since the hopping operator connects different eigenstates with the coupling parameter
\begin{equation} 
\kappa=-JFV_0\sin(\omega t)\left(\frac{d\ln J}{d\tilde{V}_0}\Bigg|_{F=0}-\frac{d\ln U}{d\tilde{V}_0}\Bigg|_{F=0}\right)\text{.}
\end{equation}
In order to identify possible excitations of the ground state $\ket{0}$ to higher-lying states $\ket{\nu}$ of the static spectrum, we look for non-vanishing matrix elements $|\matrixe{\nu}{\op{H}_J}{0}|$ of the hopping operator $\op{H}_J$.

%%%%%%%%%%%%%%%%%%%%%%%%%%%%%%%%%%%%%%%%%%%%%%%%%%%%%%%%%%%%%%%%%%%%%%%%%%%%%%%%%%%%%%%%
\begin{figure}[t]
\includegraphics{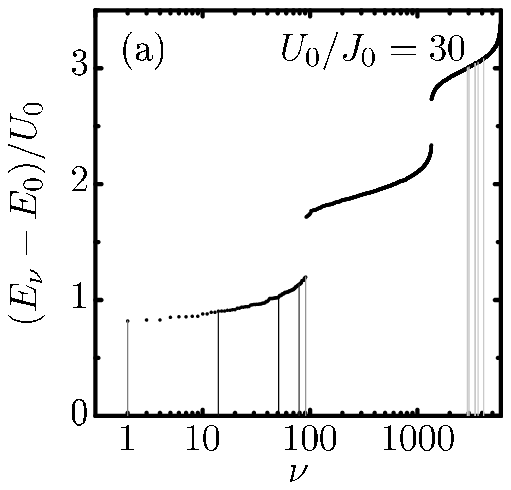}\includegraphics{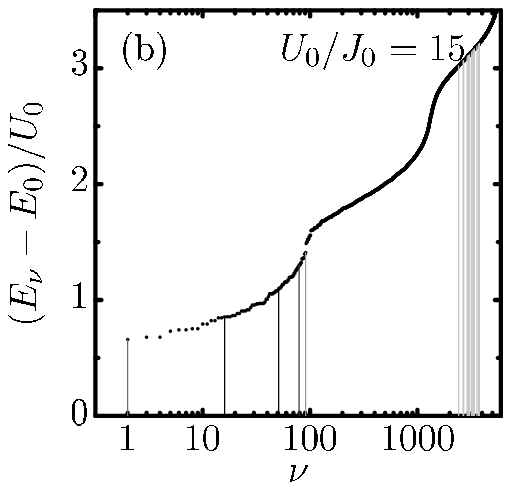}\includegraphics{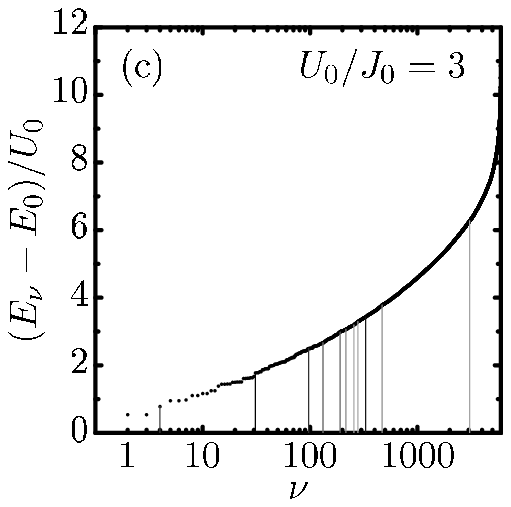}
\caption{Lowest energy eigenvalues for a system with $N=10$ bosons and $I=10$ sites at interaction strengths $U_0/J_0=30$, $15$, and $3$, computed by diagonalization of the Hamilton matrix in a truncated number basis (see text). The vertical lines mark the strongest matrix elements $|\matrixe{\nu}{\op{H}_J}{0}|$ between ground state and excited states. The strength of the matrix elements is represented by the gray level of the lines, where darker lines correspond to stronger values.}
\label{plot_spec_reg}
\end{figure}
%%%%%%%%%%%%%%%%%%%%%%%%%%%%%%%%%%%%%%%%%%%%%%%%%%%%%%%%%%%%%%%%%%%%%%%%%%%%%%%%%%%%%%%%

As a first application and a preparation for the discussion of superlattice potentials we examine a system of $10$~bosons in a regular lattice with $10$~sites. The low-lying eigenstates are obtained using a basis consisting of the $5911$ energetically lowest number states out of the complete basis ($D=92378$), generated for the interaction strength $U_0/J_0=30$ with the truncation energy $E_{\text{trunc}}/J_0=90$. Figure \ref{plot_spec_reg} shows the spectra of the truncated system for interaction strengths $U_0/J_0=30$, $15$, and $3$. 

Deep within the Mott-insulator regime, for $U_0/J_0=30$, we obtain the characteristic gapped energy spectrum shown in figure \ref{plot_spec_reg}(a). The vertical lines illustrate the strongest hopping matrix elements $|\matrixe{\nu}{\op{H}_J}{0}|$ between ground and excited states. The energy scale is shifted with respect to the ground state energy, so that the vertical lines represent excitations into the Hubbard bands ($U_0$-band) and the corresponding energies. It is remarkable that there are no sizable transition matrix elements to states in the $2U_0$-band but to the $3U_0$-band, as was pointed out in reference \cite{KoIu06,ClJa06}. With decreasing interaction strength the gaps are reduced and eventually vanish in the superfluid regime, as the sequence of plots in figure \ref{plot_spec_reg} demonstrates.

A simple interpretation of the excitations to the individual Hubbard bands can be given in the language of particle-hole excitations. In the limit of strong repulsive interactions, i.e., deep in the Mott-insulator regime, the ground state of a system with unit filling is given by $\ket{0}\approx\ket{1,1,1,1,1,1,\dots}$. The simplest mechanism to excite this state is a one-particle-one-hole excitation (1p1h) as illustrated graphically in table \ref{tab_exc}. The excitation energy associated with such a process equals the interaction strength~$\Delta E=U_0$, hence it represents a possible excitation from the ground state to a state in the first Hubbard band. Table \ref{tab_exc} lists the basic particle-hole excitations together with the Hubbard band they connect to. 

%%%%%%%%%%%%%%%%%%%%%%%%%%%%%%%%%%%%%%%%%%%%%%%%%%%%%%%%%%%%%%%%%%%%%%%%%%%%%%%%%%%%%%%%
\begin{table}[t!]
\centering
\begin{tabular}{llc}
schematic & type & energy transfer\\
\hline\hline
\\[2pt]
\includegraphics[scale=0.4]{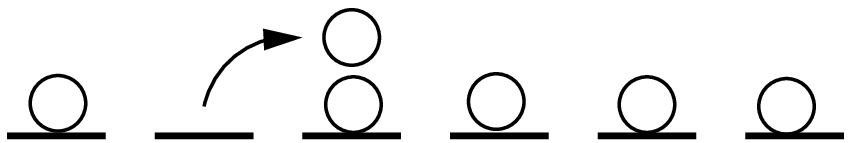} & one-particle-one-hole (1p1h) & $U_0$
\\[1pt]
\hline
\\[2pt]
\includegraphics[scale=0.4]{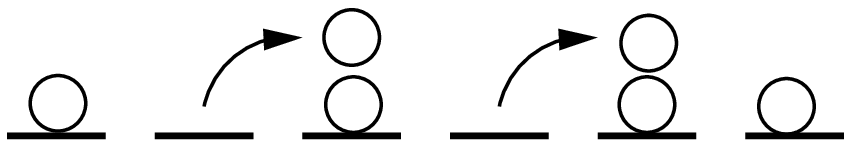} & two-particle-two-hole (2p2h) & $2U_0$
\\[1pt]
\hline
\\[2pt]
\includegraphics[scale=0.4]{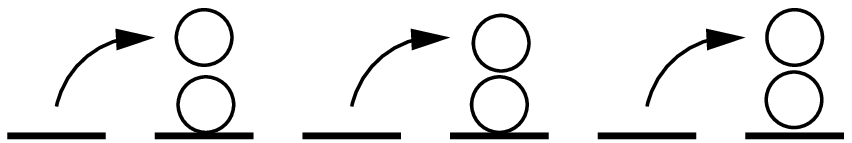} & three-particle-three-hole (3p3h) & $3U_0$
\\[1pt]
\includegraphics[scale=0.4]{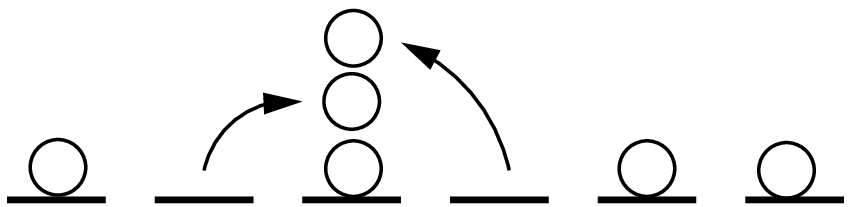} & 2p2h with two particles at same site & $3U_0$
\\[1pt]
\hline\hline	
\end{tabular}
\caption{Basic types of particle-hole excitations of first order, sorted by energy transfer.}
\label{tab_exc}
\end{table}
%%%%%%%%%%%%%%%%%%%%%%%%%%%%%%%%%%%%%%%%%%%%%%%%%%%%%%%%%%%%%%%%%%%%%%%%%%%%%%%%%%%%%%%%

%%%%%%%%%%%%%%%%%%%%%%%%%%%%%%%%%%%%%%%%%%%%%%%%%%%%%%%%%%%%%%%%%%%%%%%%%%%%%%%%%%%%%%%%
\subsection{Time evolution}

%%%%%%%%%%%%%%%%%%%%%%%%%%%%%%%%%%%%%%%%%%%%%%%%%%%%%%%%%%%%%%%%%%%%%%%%%%%%%%%%%%%%%%%%
\begin{figure}[t!]
\includegraphics{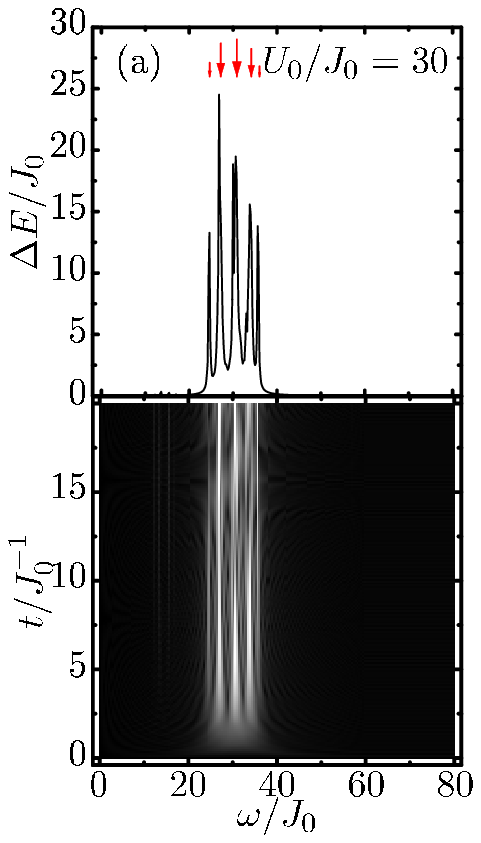}\includegraphics{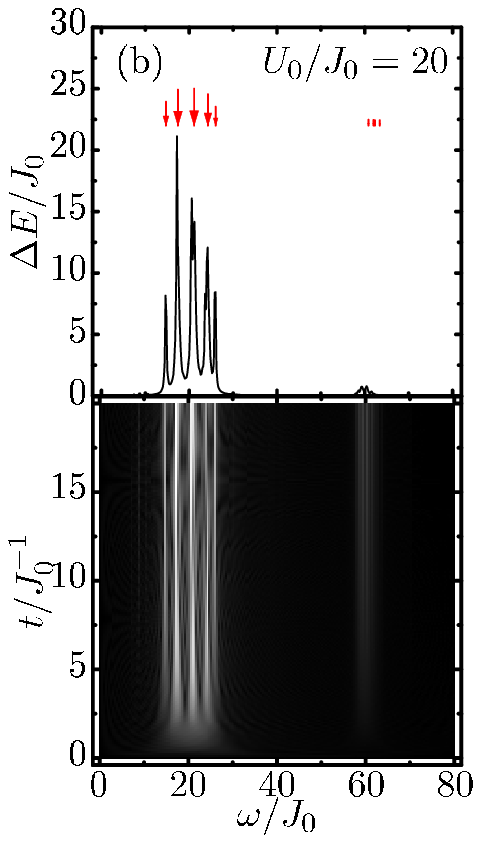}\includegraphics{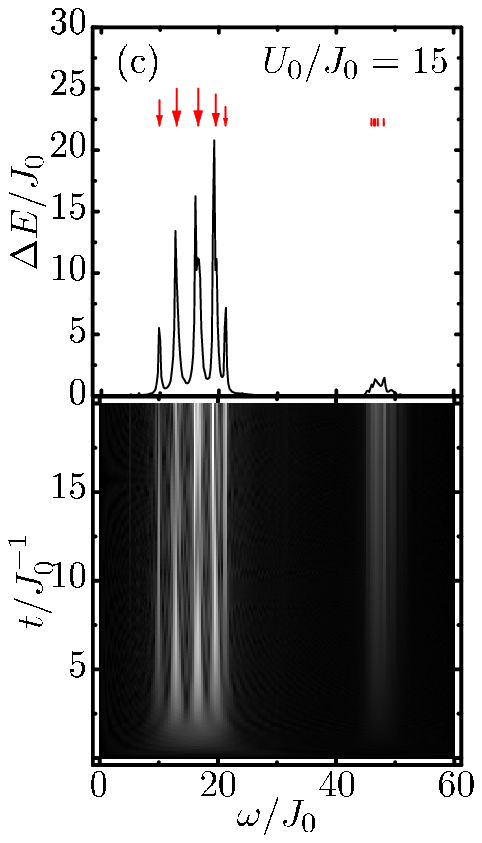}\\
\includegraphics{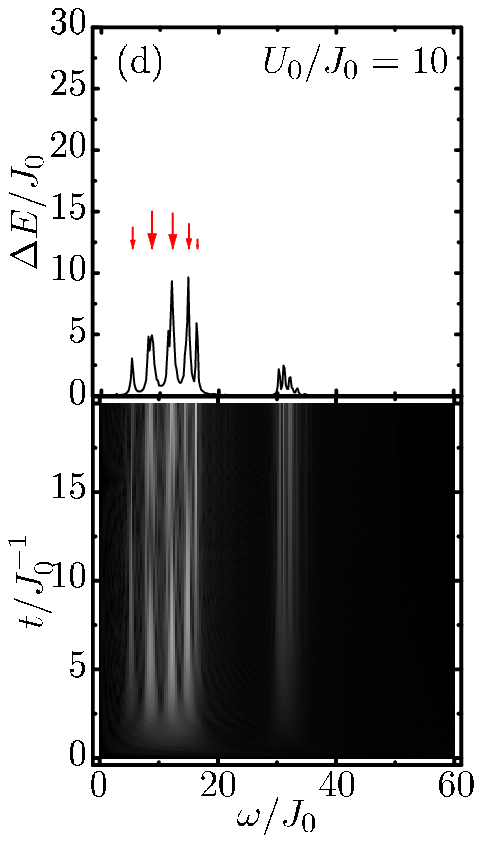}\includegraphics{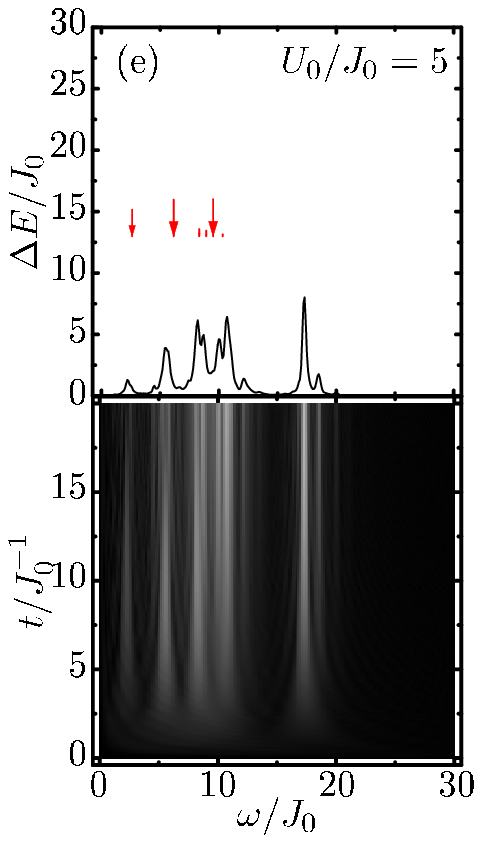}\includegraphics{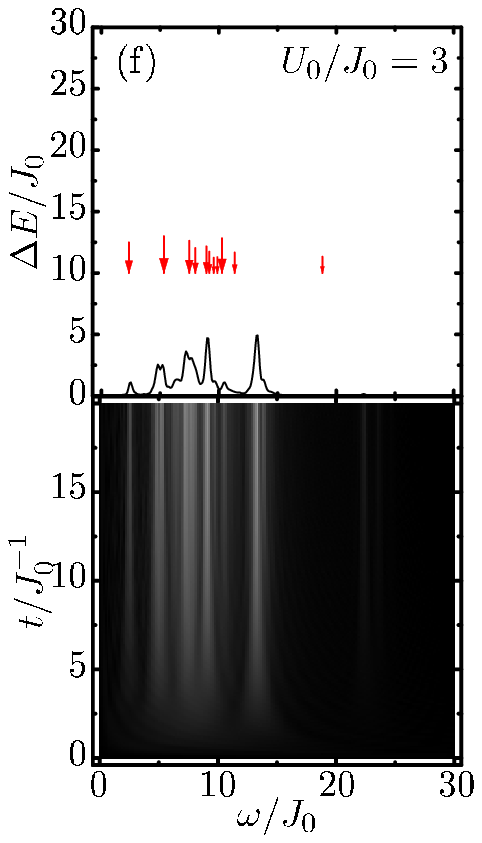}\\[-10pt]
\begin{center}\includegraphics{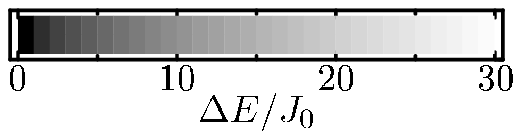}\end{center}
\caption{Energy transfer $\Delta E$ for a system of $N=10$ bosons and $I=10$ lattice sites at different interaction strengths due to the modulation of the lattice amplitude ($F=0.1$). The density plots show the energy transfer as function of modulation frequency and time, the line plots on top depict the energy transfer as function of frequency averaged over the whole evolution. The sequence from (a) to (f) corresponds to interaction strength $U_0/J_0=30$, $20$, $15$, $10$, $5$, and $3$. The (red) arrows mark the energies of eigenstates $\ket{\nu}$ which are connected to the ground state by the hopping operator $\op{H}_J$ (compare figure \ref{plot_spec_reg}).}
\label{plot_sf-mott}
\end{figure}
%%%%%%%%%%%%%%%%%%%%%%%%%%%%%%%%%%%%%%%%%%%%%%%%%%%%%%%%%%%%%%%%%%%%%%%%%%%%%%%%%%%%%%%%

In order to probe the dynamical behaviour of the Bose gas we choose a fixed ratio of the interaction to tunnelling strength $U_0/J_0$. The ground state of the system is obtained by exact diagonalization and is used as the initial state for the time evolution. Starting from this state the system is evolved in time while the lattice is modulated with frequency $\omega$ and amplitude $FV_0=0.1V_0$. The response of the system is probed via the energy transfer evaluated at each timestep using the time-evolved state $\ket{\Psi,t}$: 
\begin{equation}
\Delta E(t) = \matrixe{\Psi,t}{\op{H}_0}{\Psi,t}-E_0\text{.}
\end{equation}
$\op{H}_0$ is the Hamiltonian at time $t=0$, and $E_0$ is the ground state energy of the initial Hamiltonian.

Figure \ref{plot_sf-mott} shows the energy transfer for $10$ bosons and $10$ lattice sites at several ratios~$U_0/J_0$, ranging from the Mott insulating phase (MI), depicted in panels  \ref{plot_sf-mott}(a)-(e), to the superfluid regime (SF), shown in panel  \ref{plot_sf-mott}(f). The density plots in the lower part of each panel show the energy transfer $\Delta E$ as a function of the modulation frequency $\omega$ and time. The line plots in the upper part of each panel show the energy transfer as a function of $\omega$ and averaged over the full time evolution. The (red) arrows indicate the energies of excited states with sizable hopping matrix elements to the ground state, the size of the matrix elements is reflected by the length of the arrows.

In the simple picture of particle-hole excitations (see table \ref{tab_exc}) one would expect excitations at integer multiples of the interaction strength in the strongly correlated regime. The simulations illustrated in figure \ref{plot_sf-mott}(a)-(d) confirm this assumption partially. A large resonance appears at the frequency $\omega=U_0$ in the Mott phase, in panels (b)-(d) weaker excitations also appear at $\omega=3U_0$. In the case of the simulation for $U_0/J_0=30$ the truncated basis does not include number states with energies of $2U_0$ and higher, and consequently the $3U_0$-resonance does not appear.

%%%%%%%%%%%%%%%%%%%%%%%%%%%%%%%%%%%%%%%%%%%%%%%%%%%%%%%%%%%%%%%%%%%%%%%%%%%%%%%%%%%%%%%%
\begin{figure}[t]
\includegraphics{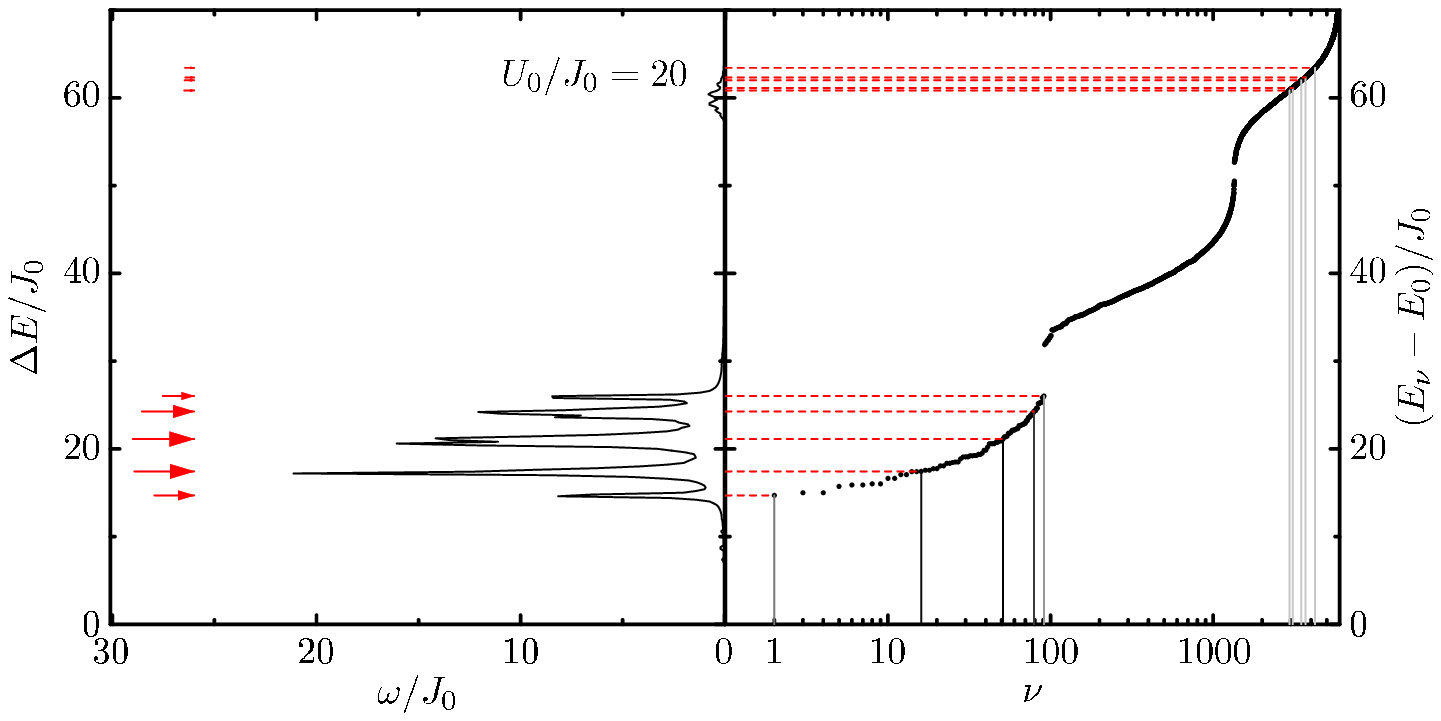}
\caption{Relation between the excitation energies of the static Hubbard Hamiltonian and the resonance structure of a bosonic system ($I=N=10$) at interaction strength $U_0/J_0=20$. The left panel shows the time-averaged energy transfer as a function of the frequency $\omega$; the right panel depicts the lower part of the eigenvalue spectrum of the corresponding static Hamiltonian. The vertical gray lines mark the eigenstates which are connected to the ground state by strong matrix elements $|\matrixe{\nu}{\op{H}_J}{0}|$. The horizontal (red) lines point out the correspondence between the excitation energies and the resonance structure.}
\label{plot_substruct}
\end{figure}
%%%%%%%%%%%%%%%%%%%%%%%%%%%%%%%%%%%%%%%%%%%%%%%%%%%%%%%%%%%%%%%%%%%%%%%%%%%%%%%%%%%%%%%%

The absence of a resonance peak at frequency $\omega=2U_0$ was already evident from the linear response analysis illustrated in figure \ref{plot_spec_reg}(a)-(c), which shows no significant matrix elements between the ground state and excited states in the $2U_0$-band. This reveals that 2p2h-processes as schematically depicted in table \ref{tab_exc} are not relevant for the excitation.

Towards the superfluid regime the resonances are significantly broadened, which is in agreement with experimental evidence \cite{StMo04}. Figure \ref{plot_sf-mott}(e) shows a simulation in the transition region from Mott-insulator to superfluid around $(U_0/J_0)_c=4.65$ \cite{RoBu03a}. The response spectrum reveals a significant shift of the resonance towards higher frequencies for decreasing interaction strength. The simple approximation of a single number state as the ground state is not adequate for weaker interactions and especially in the superfluid regime due to the delocalization of the atoms. The admixture of other number states leads to additional particle-hole excitations and therefore to a broadening of the resonances.

The fine-structure of the resonances (figure \ref{plot_spec_reg}) can be explained by the linear response analysis. Figure \ref{plot_substruct} reveals the connection between the excitations to higher eigenstates and the structure of the resonances appearing in the time evolution. The left hand side shows the time-averaged energy transfer of a $I=N=10$ bosonic system, the right hand side the corresponding energy spectrum obtained by solving the eigenvalue problem of the initial Hamiltonian. The vertical lines indicate the eigenstates that are connected to the ground state by large matrix elements of the hopping operator~$H_J$. The horizontal dashed (red) lines and the (red) arrows mark the corresponding excitation energies, which resemble --- with remarkable precision --- the fine-structure of the resonances emerging in the time evolution.  

In addition to the resonances at integer multiples of the interaction strength, there is also a tiny resonance at $U_0/2$ in the strongly interacting cases. It results from non-linear effects, such as the absorption of two photons of the energy $U_0/2$, which are not captured by the linear response analysis.

%%%%%%%%%%%%%%%%%%%%%%%%%%%%%%%%%%%%%%%%%%%%%%%%%%%%%%%%%%%%%%%%%%%%%%%%%%%%%%%%%%%%%%%%%%%%%%%%
\subsection{Benchmark for the truncation scheme}

In section \ref{sec_bhm} we introduced the basis truncation scheme in order to reduce the numerical effort of the simulations. For static ground state properties, we have shown that the basis dimension can be reduced typically by two orders of magnitude without a significant loss of precision \cite{ScHi06}. This does not necessarily hold for the exact time evolution discussed in this paper. In order to describe the excitation effects properly, one has to use a basis which allows to describe more than a few eigenstates.

%%%%%%%%%%%%%%%%%%%%%%%%%%%%%%%%%%%%%%%%%%%%%%%%%%%%%%%%%%%%%%%%%%%%%%%%%%%%%%%%%%%%%%%%
\begin{figure}[t]
\includegraphics{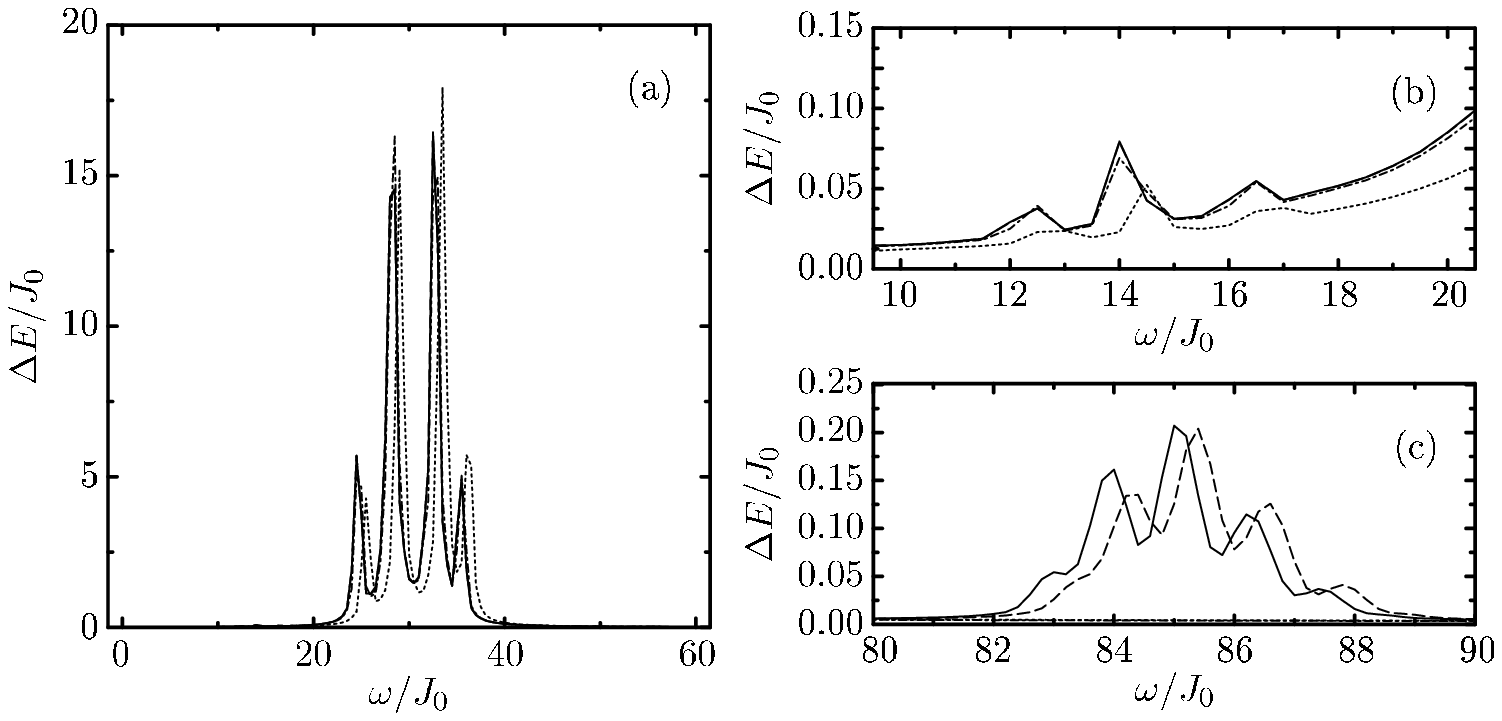}
\caption{Comparison of the time-averaged energy transfer for a Bose gas with $I=N=8$ at the interaction strength $U_0/J_0=30$. The different lines show the result for the full basis \linemediumsolid\ and for truncated bases with  $E_{\text{trunc}}/J_0=90$ \linemediumdashed, $E_{\text{trunc}}/J_0=60$ \linemediumdashdot, and $E_{\text{trunc}}/J_0=30$ \linemediumdotted. Depicted are the main resonance at $\omega=U_0$ (a), the non-linear resonance at $\omega=U_0/2$ (b), and the $3U_0$-resonance (c).}
\label{plot_trunc-test}
\end{figure}
%%%%%%%%%%%%%%%%%%%%%%%%%%%%%%%%%%%%%%%%%%%%%%%%%%%%%%%%%%%%%%%%%%%%%%%%%%%%%%%%%%%%%%%%

To test the basis truncation in the context of time-dependent calculations with a modulated lattice we compare time evolutions with several cut-off energies $E_{\text{trunc}}$. We perform the time evolution of a system of $N=8$~boson in $I=8$~sites for $U_0/J_0=30$, which is the largest system with integer filling that allows the use of the complete basis. We compare the energy transfer as  a function of the modulation frequency $\omega$, averaged over the evolution time $t_{\max}/J_0^{-1}=10$. Table \ref{tab_trunc} specifies the characteristics of the bases used for the comparison. Figure~\ref{plot_trunc-test}(a) shows the results for the $U_0$-resonance. The response obtained with the basis truncated at $E_{\text{trunc}}/J_0=90$ is in excellent agreement with the calculation using the complete basis. The basis with cut-off energy $E_{\text{trunc}}/J_0=60$ shows small deviations in the peaks. The basis truncated at $E_{\text{trunc}}/J_0=30$ reproduces the peak as well as its fine-structure but the whole resonance is shifted by roughly $\Delta E=1.5J_0$. The shift towards higher energies for more severe truncations can be explained by the Hylleraas-Undheim-MacDonald theorem \cite{HyUn30,Ma33}, which states that the exact eigenenergies are lower bounds for the corresponding eigenvalues obtained with a truncated basis. In a variational picture, the basis truncation reduces the flexibility of the number state representation (trial states) and thus leads to an increase of the energy eigenvalues in the truncated space. 

Figure~\ref{plot_trunc-test}(b) shows an enlarged energy interval around the non-linear resonance at frequency~$\omega=U_0/2$. The basis with the truncation energy $E_{\text{trunc}}/J_0=90$ perfectly reproduces the result of the complete basis, even the $E_{\text{trunc}}/J_0=60$-basis causes only small deviations. In agreement with the results for the $U_0$-resonance, the whole resonance is significantly shifted to higher energies in the simulation using the $E_{\text{trunc}}/J_0=30$-basis due to the overestimation of the eigenenergies. A closer look at the $3U_0$-resonance in figure~\ref{plot_trunc-test}(c) illustrates the limitations of truncated bases for the description of high-lying resonances. For truncation energies $E_{\text{trunc}}/J_0=30$ and $60$ the bases do not include any number states that correspond to energies as high as the excitation energy, hence no energy transfer is possible. The basis with the truncation energy $E_{\text{trunc}}/J_0=90$ just about includes the numbers states matching the excitation energy of $\omega=3U_0=90J_0$, and consequently, an energy shift occurs, which is comparable in size to the shift of the $U_0$-resonance for the truncated basis with $E_{\text{trunc}}/J_0=30$. 

%%%%%%%%%%%%%%%%%%%%%%%%%%%%%%%%%%%%%%%%%%%%%%%%%%%%%%%%%%%%%%%%%%%%%%%%%%%%%%%%%%%%%%%%
\begin{table}[t!]
\centering
\begin{tabular}{cll}
$E_{\text{trunc}}/J_0$ & basis dimension & excitations included \\
\hline\hline\\[-10pt]
$\infty$	 	& 6435 (complete)	& all	\\
90		& 1205	& up to 3-particle-3-hole  \\	
60		& 477 	& up to 2-particle-2-hole  \\
30		& 57 	& up to 1-particle-1-hole  \\
\hline\hline
\end{tabular}
\caption{Main characteristics of the bases used for the comparison. Shown are the truncation energy $E_{\text{trunc}}$, the basis dimension and the highest included particle-hole excitation with respect to the reference state $\ket{1,1,1,1,1,1,1,1}$. }
\label{tab_trunc}
\end{table}
%%%%%%%%%%%%%%%%%%%%%%%%%%%%%%%%%%%%%%%%%%%%%%%%%%%%%%%%%%%%%%%%%%%%%%%%%%%%%%%%%%%%%%%%

%%%%%%%%%%%%%%%%%%%%%%%%%%%%%%%%%%%%%%%%%%%%%%%%%%%%%%%%%%%%%%%%%%%%%%%%%%%%%%%%%%%%%%%%
\section{Bose gases in a two-colour superlattice}\label{sec_superlattice}

\subsection{Two-colour superlattice and the phase diagram}

%%%%%%%%%%%%%%%%%%%%%%%%%%%%%%%%%%%%%%%%%%%%%%%%%%%%%%%%%%%%%%%%%%%%%%%%%%%%%%%%%%%%%%%%
\begin{figure}[t!]
\includegraphics{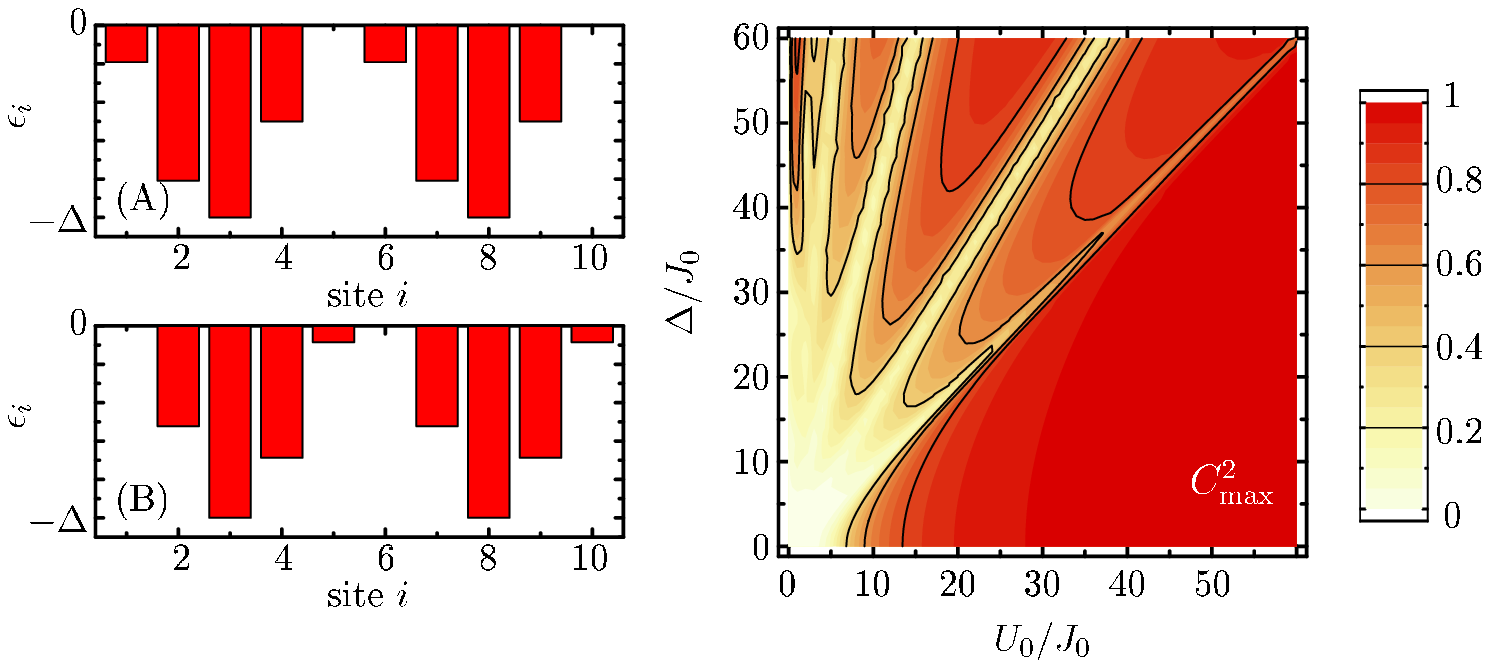}
\caption{Left panels: Two different sets of on-site energies $\epsilon_{i}$ of period-five two-colour superlattices used in this paper. Right panel: Phase diagram of bosons ($N=10$) in an optical superlattice ($I=10$) spanned by the interaction strength~$U_0/J_0$ and the superlattice amplitude $\Delta/J_0$ for the set (A) of on-site energies. Depicted is the maximum coefficient $C^2_{\text{max}}$ of the number state expansion (\ref{eq_eigenstate}) of the ground state.}
\label{plot_phase_diag}
\end{figure}
%%%%%%%%%%%%%%%%%%%%%%%%%%%%%%%%%%%%%%%%%%%%%%%%%%%%%%%%%%%%%%%%%%%%%%%%%%%%%%%%%%%%%%%%

In this section the dynamical signatures of Bose gases in two-colour superlattices are investigated. In experiment, superlattices can be generated by a superposition of two optical standing waves, which leads to a sinusoidal modulation of the depth of the individual lattice wells. The spatial modulation enters the Hubbard model via the external potential term discussed in section \ref{sec_bhm}, with an appropriate distribution of the on-site energies $\epsilon_{i}$. The left hand side of figure \ref{plot_phase_diag} shows two different distributions of on-site energies for a two-cell superlattice of 10 sites. Essentially, the two superlattices differ in the relative phase of the standing waves. We will use these two realizations to estimate the dependence of the response on the detailed topology of the superlattice. In the following we refer to these on-site energy distributions as superlattices (A) and (B).

The amplitude $\Delta$ of the superlattice is an additional parameter which generates a rich structure in the phase diagram \cite{RoBu03b,RoBu03c}. The right hand side of figure \ref{plot_phase_diag} shows the two-dimensional phase diagram for $N=10$ bosons and $I=10$ sites with on-site energies $\epsilon_{i}$ according to superlattice (A) (figure \ref{plot_phase_diag}). Plotted is the maximum coefficient $C^2_{\text{max}}$ of the expansion (\ref{eq_eigenstate}) of the ground state in the number basis. The darker (red) shadings represent regions of a large maximum coefficient which indicates that a single number state is dominating the ground state. Brighter shadings refer to small maximum coefficients which correspond to a ground state given by a superposition of many number states.

%%%%%%%%%%%%%%%%%%%%%%%%%%%%%%%%%%%%%%%%%%%%%%%%%%%%%%%%%%%%%%%%%%%%%%%%%%%%%%%%%%%%%%%%
\begin{figure}[t!]
\centering
\psfrag{Delta}{$\Delta$}
\psfrag{label-a}{(a)}
\psfrag{label-b}{(b)}
\includegraphics[scale=0.5]{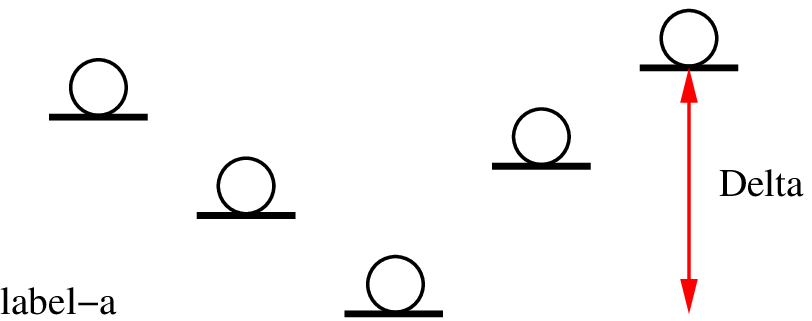}\hspace{2cm}
\includegraphics[scale=0.5]{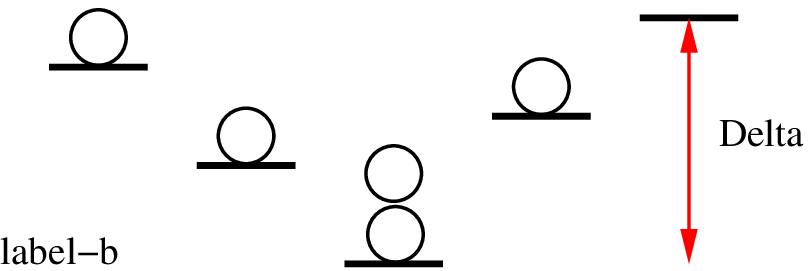}
\caption{Schematic diagram of a single cell of the superlattice: If the superlattice amplitude $\Delta$ is close to the interaction strength $U_0$, the number states (a) and (b) have roughly the same energy. The relative height between the sites represents the different on-site energies $\epsilon_i$.}
\label{car_bg}
\end{figure}
%%%%%%%%%%%%%%%%%%%%%%%%%%%%%%%%%%%%%%%%%%%%%%%%%%%%%%%%%%%%%%%%%%%%%%%%%%%%%%%%%%%%%%%%

The bright region at small interaction strengths and superlattice amplitudes indicates the superfluid phase. By increasing the interaction strength the system enters the homogenous Mott-insulator phase, which is represented by the dark (red) area below the diagonal~($U_0=\Delta$). As long as the interaction strength exceeds the superlattice amplitude it is energetically unfavourable to occupy a site with more than one boson. Close to the point where amplitude and interaction are equal, the maximum coefficient $C^2_{\text{max}}$ decreases: Here, the two number states depicted in figure \ref{car_bg} have the same energy, since the energy gain in on-site energy obtained by hopping to the deepest well (figure \ref{car_bg}(b)) is compensated by the increase in interaction energy. 

These redistributions of the particles in favour of the deep lattice wells become more and more important if the superlattice amplitude increases further. As soon as the gain in on-site energy for a particular redistribution exceeds the loss due to the increased interaction energy, the particle will move to a deeper lattice well. These successive redistributions generate the lobe structure in the phase diagram (figure \ref{plot_phase_diag}). In the case of random lattices the redistributions happen almost continuously, giving rise to the Bose glass phase. Eventually all atoms are localized at the deepest lattice well. This complete localization also appears for small or vanishing interaction strengths $U_0/J_0$ in the phase diagram.

%%%%%%%%%%%%%%%%%%%%%%%%%%%%%%%%%%%%%%%%%%%%%%%%%%%%%%%%%%%%%%%%%%%%%%%%%%%%%%%%%%%%%%%%
\subsection{Linear response analysis}

%%%%%%%%%%%%%%%%%%%%%%%%%%%%%%%%%%%%%%%%%%%%%%%%%%%%%%%%%%%%%%%%%%%%%%%%%%%%%%%%%%%%%%%%
\begin{figure}[t!]
\includegraphics{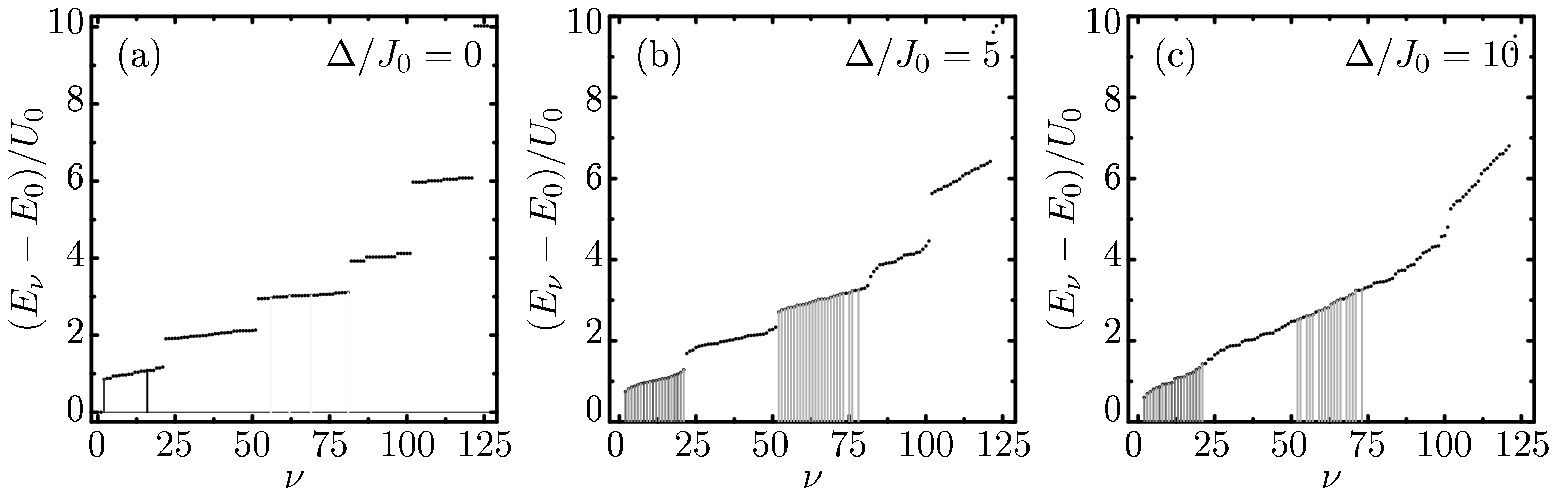}
\includegraphics{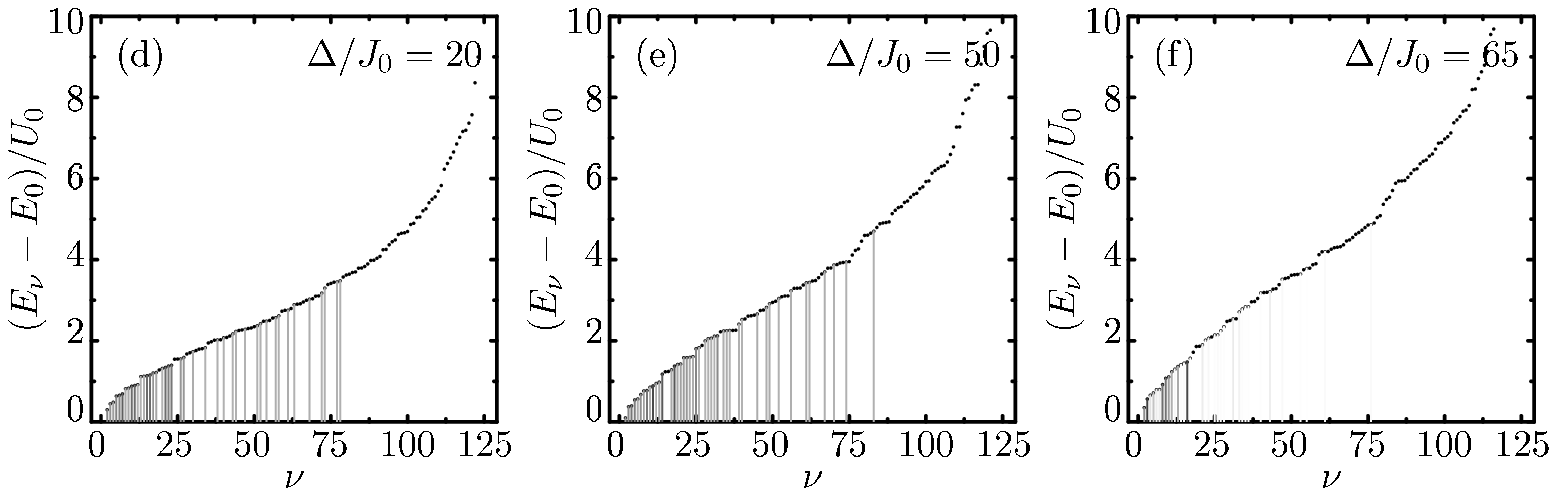}
\caption{Energy spectrum of an $I=N=5$ system at fixed interaction strength $U_0/J_0=30$ and various superlattice amplitudes $\Delta/J_0$. The vertical lines mark the eigenstates that are connected to the ground state via the hopping operator $\op{H}_J$.}
\label{plot_spect}
\end{figure}
%%%%%%%%%%%%%%%%%%%%%%%%%%%%%%%%%%%%%%%%%%%%%%%%%%%%%%%%%%%%%%%%%%%%%%%%%%%%%%%%%%%%%%%%

As a first analysis of the response of superlattice systems we examine the spectrum of the static Hamiltonian (\ref{eq_hamop}) for a single superlattice cell. Figure \ref{plot_spect} shows the spectra of the Bose-Hubbard Hamiltonian for $N=5$~bosons and $I=5$~sites at fixed interaction strength~$U_0/J_0=30$. Instead of two-cell superlattices such as (A) and (B) we consider a single cell only, because the small basis dimensions allow us to solve the full eigenproblem without truncation. Note that the energies plotted in figure \ref{plot_spect} are shifted by the ground state energy in such a manner that the ground state energy is zero.

The sequence of plots in figure \ref{plot_spect} reveals the effect of the superlattice amplitude~$\Delta$ on the Hubbard band structure. The vertical lines mark the eigenstates which are connected to the ground state via nonvanishing matrix elements of the hopping operator~$\op{H}_J$. The relative strength 
of these matrix elements $|\matrixe{\nu}{\op{H}_J}{0}|$ is indicated by the gray level of the lines. Darker shadings correspond to larger matrix elements. 

In the absence of a spatial modulation the typical Hubbard bands of the strongly correlated regime are visible, as depicted in figure \ref{plot_spect}(a). As discussed in section \ref{sec_reg_lattice}, the second Hubbard band ($2U_0$-band) is not connected to the ground state in first order. Figure~\ref{plot_spect}(b) shows the eigenenergy spectrum for a system with a small superlattice amplitude of $\Delta/J_0=5$. The gapped structure is still visible but the width of the individual bands is significantly increased and the gaps are reduced. The breaking of the spatial symmetry by the superlattice generates a larger number of matrix elements which connect the ground state to higher bands. At a superlattice amplitude $\Delta/J_0=10$, shown in figure \ref{plot_spect}(c), the gaps have completely vanished, but nevertheless, the largest matrix elements still couple the ground state to eigenstates in the $U_0$ and $3U_0$ bands. 

At an amplitude $\Delta/J_0=20$, which is still within the homogenous Mott-insulator phase, the separation  of the sizable matrix elements which were initially associated with the $U_0$ and $3U_0$ bands dissolves as shown in figure \ref{plot_spect}(d). An interval of large matrix elements remains for excitation energies up to $1.5 U_0$, which resembles the original $U_0$-band. Beyond that, there is a number of weaker matrix elements covering excitation energies of up to $3.5 U_0$. With increasing superlattice amplitude, any semblance of the band structure disappears. As depicted in figure \ref{plot_spect}(e) for $\Delta/J_0=50$ one finds a continuous distribution of matrix elements with dominant matrix elements at low excitation energies. For very large superlattice amplitudes, e.g. for $\Delta \approx 2 U_0$ as depicted in figure \ref{plot_spect}(f), the higher-lying matrix elements are suppressed and strong matrix elements appear only at the lower end of the energy scale. This characteristic behaviour emerges also from the dynamical simulations.

%%%%%%%%%%%%%%%%%%%%%%%%%%%%%%%%%%%%%%%%%%%%%%%%%%%%%%%%%%%%%%%%%%%%%%%%%%%%%%%%%%%%%%%%
\subsection{Time evolution}

%%%%%%%%%%%%%%%%%%%%%%%%%%%%%%%%%%%%%%%%%%%%%%%%%%%%%%%%%%%%%%%%%%%%%%%%%%%%%%%%%%%%%%%%
\begin{figure}[t]
\begin{minipage}{0.66\linewidth}
\includegraphics{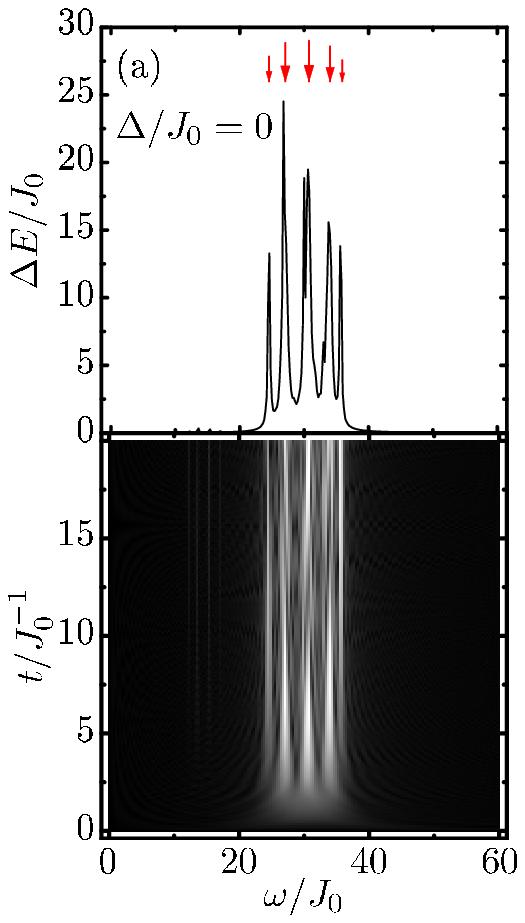}\includegraphics{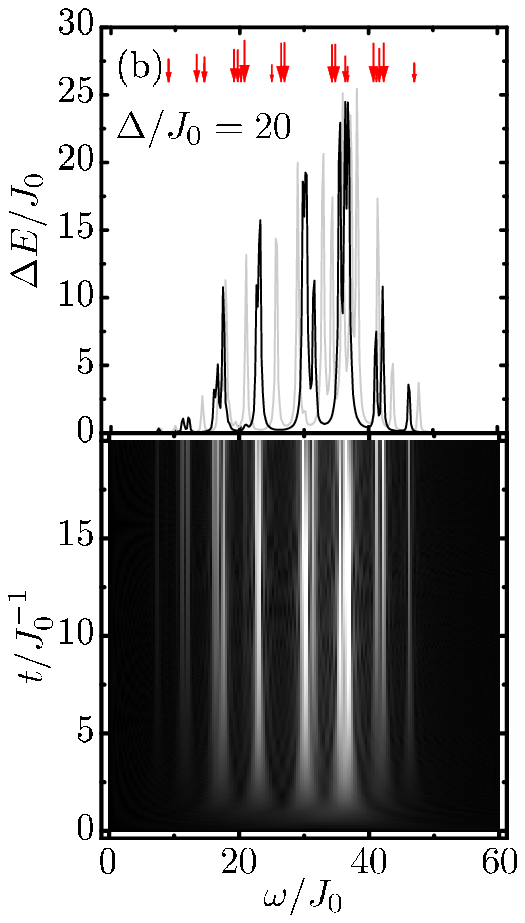}
\end{minipage}
\hspace{10mm}\begin{minipage}{0.33\linewidth}
\includegraphics{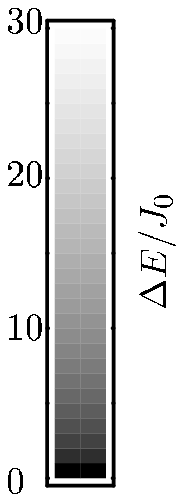}
\end{minipage}
\caption{Energy transfer during lattice modulation ($F=0.1$) for a superlattice system with $I=N=10$ and interaction strength~$U_0/J_0=30$ for two superlattice amplitudes~$\Delta<U_0$ in the homogenous Mott-insulator phase. The line plots show the time-averaged energy transfer as function of $\omega$ for the superlattices (A) \linemediumsolid\ and (B) \linemediumsolid[lgray]\ defined in figure \ref{car_bg}.
The (red) arrows mark the excitation energies predicted by linear response analysis. The density plots illustrate the energy transfer as a function of frequency and time for the superlattice (A).}
\label{plot_superlat1}
\end{figure}
%%%%%%%%%%%%%%%%%%%%%%%%%%%%%%%%%%%%%%%%%%%%%%%%%%%%%%%%%%%%%%%%%%%%%%%%%%%%%%%%%%%%%%%%

%%%%%%%%%%%%%%%%%%%%%%%%%%%%%%%%%%%%%%%%%%%%%%%%%%%%%%%%%%%%%%%%%%%%%%%%%%%%%%%%%%%%%%%%
\begin{figure}[t]
\includegraphics{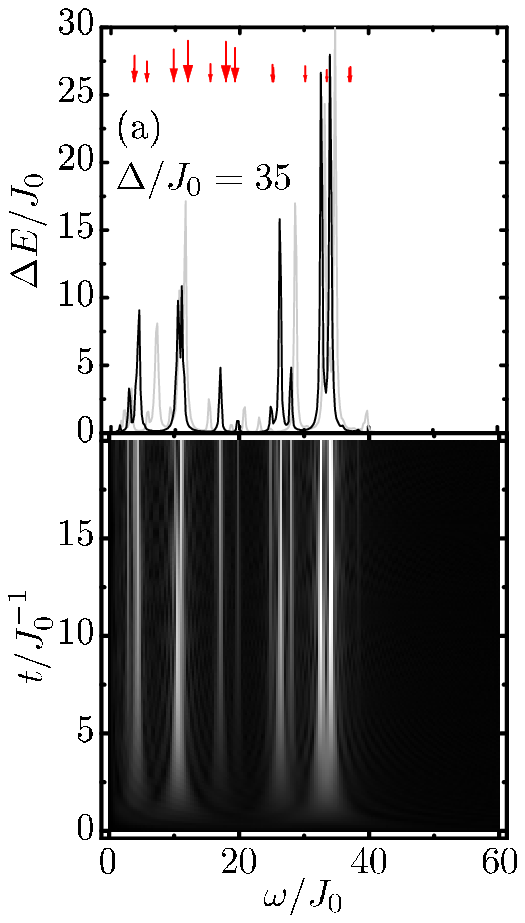}\includegraphics{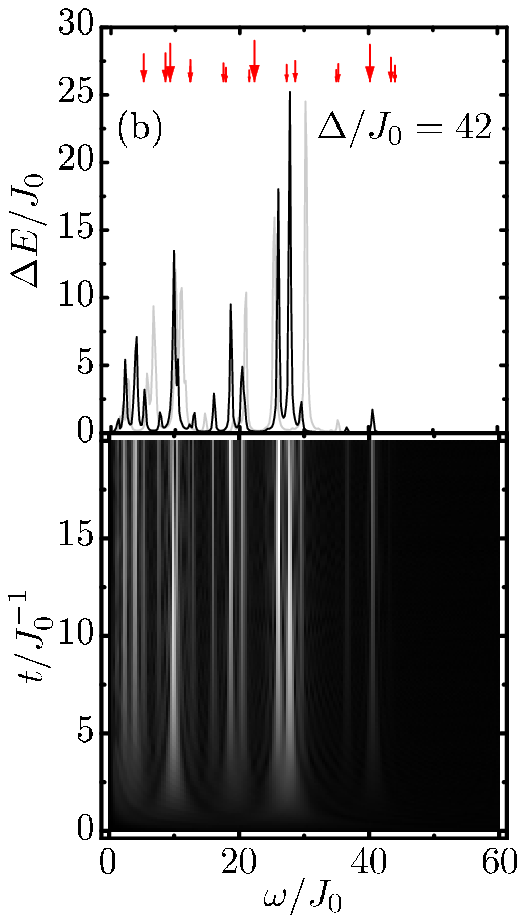}\includegraphics{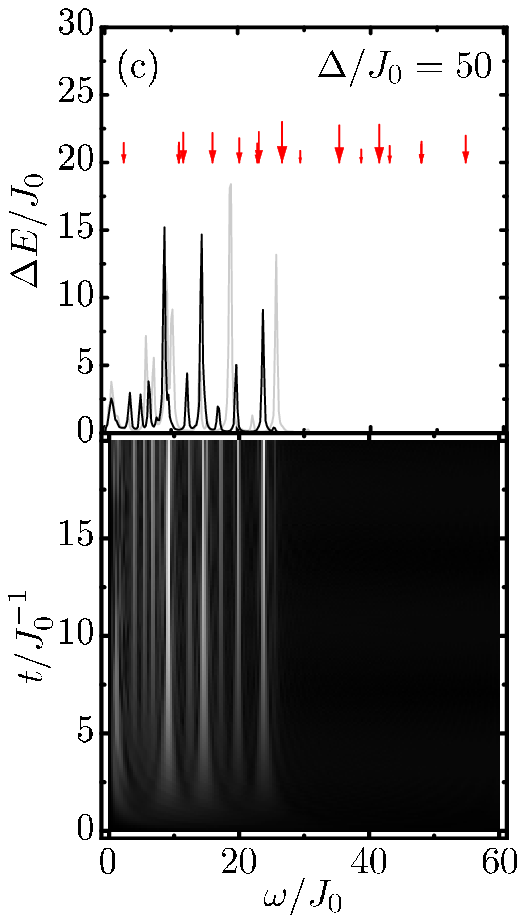}\\[-8pt]
\begin{center}\includegraphics{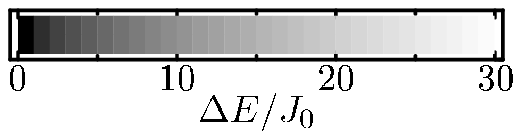}\end{center}
\vspace{-12pt}
\caption{Energy transfer during lattice modulation ($F=0.1$) for a superlattice system with $I=N=10$ and interaction strength~$U_0/J_0=30$ for three superlattice amplitudes~$\Delta>U_0$ in the quasi Bose-glass phase. The line plots show the time-averaged energy transfer as function of $\omega$ for the superlattices (A) \linemediumsolid\ and (B) \linemediumsolid[lgray]\ defined in figure \ref{car_bg}.
The (red) arrows mark the excitation energies predicted by linear response analysis. The density plots illustrate the energy transfer as a function of frequency and time for the superlattice (A).}
\label{plot_superlat2}
\end{figure}
%%%%%%%%%%%%%%%%%%%%%%%%%%%%%%%%%%%%%%%%%%%%%%%%%%%%%%%%%%%%%%%%%%%%%%%%%%%%%%%%%%%%%%%%

We now turn to fully dynamical simulations of Bose gases in time-dependent superlattices. We focus on a system of $N=10$ bosons in the two-cell superlattices with $I=10$ sites as defined by (A) and (B) in figure \ref{plot_phase_diag}. All simulations are performed for a fixed interaction strength $U_0/J_0=30$ and various values for the superlattice amplitude $\Delta/J_0$. As initial state we use the ground state of the static Bose-Hubbard Hamiltonian for the same parameters. In analogy to the procedure for regular potentials in section \ref{sec_reg_lattice} the lattice potential is modulated in time with a frequency $\omega$ and a fixed relative amplitude $F=0.1$. 

Figures \ref{plot_superlat1} and \ref{plot_superlat2} show the energy transfer to the system at several
superlattice amplitudes for a fixed interaction of $U_0/J_0=30$. All simulations are performed with each of the two on-site energy distributions depicted on the left of figure \ref{plot_phase_diag} in order to assess the impact of the detailed distribution of the on-site energies $\epsilon_i$. The density plots in the lower part of each panel show the energy transfer as function of time and frequency for superlattice (A). The plots in the upper part represent the time-averaged energy transfer as function of the frequency for both superlattice topologies. The (red) arrows above the individual peaks mark the excitation energies associated with the strongest matrixelements $|\matrixe{\nu}{\op{H}_J}{0}|$. The size of the arrows indicates the relative strength of the matrixelements.

Figure \ref{plot_superlat1}(a) depicts the $U_0$-resonance in the case of a regular lattice, i.e., for vanishing superlattice amplitude $\Delta/J_0$. As pointed out before, the excitation energies associated with the strongest matrix elements resulting from the linear response analysis nicely describe the position and fine-structure of the resonance. If we increase the superlattice amplitude to the value $\Delta/J_0=20$ --- still remaining in the homogeneous Mott-insulator phase --- the overall width of the resonance structure increases. The characteristic scale of the fine structure increases as well, and additional peaks emerge. This is in accord with the results of the linear response analysis of the previous section. An increase of the superlattice amplitude leads to a broadening of the Hubbard bands (cf. figure \ref{plot_spect} (b)) which corresponds to the broadening of the resonance. The larger number of peaks in the resonance corresponds to the larger number of matrix elements. The comparison of the two different superlattice topologies shows that the envelopes of the resonance are practically identical in both cases. Only the details of the fine structure depend on the particular set of on-site energies $\epsilon_i$ used.

After crossing the transition to the quasi Bose-glass phase at $\Delta=U_0$, the resonance structure changes dramatically as shown in figure \ref{plot_superlat2}. Already at $\Delta/J_0=35$, i.e., slightly above the transition at $\Delta/J_0=U_0/J_0=30$, the resonance is completely fragmented. There is no longer a smooth envelope with centroid at $\omega=U_0$. Instead, the strength is split into two groups of individual peaks: one group around the original resonance position and another group at low excitation energies. In particular, we observe low-lying resonances at $\omega/J_0<10$ --- a regime where no response was observed for a system in the homogeneous Mott-insulator phase. With increasing superlattice amplitude, the response is shifted towards lower excitation energies, as indicated by the sequence of plots in figure \ref{plot_superlat2}. At $\Delta/J_0=50$, for instance, no response is left at the original resonance position $\omega=U_0$ and all peaks are concentrated at low excitation energies. 

This characteristic behaviour of the response appears to be a clear signature for the Mott-insulator to quasi Bose-glass transition for boson in superlattices, which is directly accessible to experiments. The comparison of the two different superlattice topologies, both with a period of five sites, demonstrates that the fine-structure of the response depends on the details of the superlattice, but that the gross characteristics are not affected.  

Finally, one should note that the linear response analysis already hints at these substantial changes in the resonance spectrum. For larger superlattice amplitudes however, effects beyond the simple linear perturbation scheme become increasingly important and lead to significant discrepancies in comparison to the full time-dependent simulation (cf. figure \ref{plot_superlat2}(c)).

%%%%%%%%%%%%%%%%%%%%%%%%%%%%%%%%%%%%%%%%%%%%%%%%%%%%%%%%%%%%%%%%%%%%%%%%%%%%%%%%%%%%%%%%%%%%%%%%%
\section{Summary and conclusions}

We have studied the dynamics of Bose gases in one dimensional optical lattices and superlattices with time dependent lattice amplitudes. The time evolution based on the time-dependent Bose-Hubbard Hamiltonian is performed numerically for systems with up to 10 bosons and 10 lattice sites. In order to reduce the numerical effort we have used an \textit{a priori} basis truncation scheme \cite{ScHi06}, which reduces the dimension of the number basis to a tractable size. The control parameter of this truncation scheme is the cut-off energy $E_{\text{trunc}}$, which defines the maximum energy of the number states in the basis. We have shown that the truncation also allows reliable dynamical calculations.

As a first application, we have examined Bose gases in a regular lattice potential. In agreement with experiment \cite{StMo04} and other theoretical results \cite{KBM05,KoIu06,ClJa06}, we illustrated the characteristic resonance structure of the Bose gas in the strongly correlated Mott-insulator regime, which is washed out and broadened towards the superfluid phase. 

Treating the time-dependence of the Bose-Hubbard Hamiltonian as a linear perturbation demonstrates that excitations are generated by the hopping operator in first order~\cite{IuCa06,KoIu06,ClJa06}. The analysis of matrix elements of the hopping operator between ground and excited states allows a prediction of the resonance position and a detailed explanation of the fine structure. It is shown, that the individual peaks of a resonance correspond to large matrix elements $|\matrixe{0}{\op{H}_J}{\nu}|$, where $\ket{\nu}$ are the eigenstates of the initial Hamiltonian. 

In the second part we have investigated dynamical signatures of quantum phase transitions in two-colour superlattice potentials. We have shown that for superlattice amplitudes $\Delta$ smaller than the interaction strength $U_0$, i.e., in the homogeneous Mott insulator phase, the resonances resemble those of regular lattices. With increasing superlattice amplitude, the resonance structures become broader and additional peaks appear in their fine-structure. This finding is consistent with recent experiments \cite{FaLy06} using time-dependent superlattices with incommensurate wavelengths of the superimposed standing waves. First calculations for incommensurate superlattices confirm that the gross characteristics of the resonance spectrum are independent of the precise lattice topology, only the fine-structure of the resonances is affected.  

As soon as the superlattice amplitude exceeds the interaction strength, i.e., on entering the quasi Bose-glass phase, the resonance structure changes completely. The resonance at $\omega=U_0$ is fragmented and a strong low-energy component develops. Further increase of the superlattice amplitude eventually suppresses all strengths at the original resonance position, such that only the response at very low excitation energies remains. This characteristic behaviour might serve as an experimental indicator for the transition from homogeneous Mott-insulator to quasi Bose-glass phase.

In order to address the experimental scenario \cite{FaLy06} directly, we are going to investigate the influence of different lattice topologies, external trapping potentials, and filling factors in detail \cite{HiSc07}. Nevertheless, the results presented in this paper apply to the Mott domain with one atom per site even in the presence of a weak trapping potential. Another topic for future studies is the impact of non-zero temperatures on the response behaviour, which is beyond the present zero-temperature formalism.

%%%%%%%%%%%%%%%%%%%%%%%%%%%%%%%%%%%%%%%%%%%%%%%%%%%%%%%%%%%%%%%%%%%%%%%%%%%%%%%%%%%%%%%%%
\section*{Acknowledgements}
We thank K. Braun-Munzinger, J. Dunningham, and K. Burnett for fruitful discussions and for providing reference data during the development of the time-evolution code. RR thanks the DFG for support.

%%%%%%%%%%%%%%%%%%%%%%%%%%%%%%%%%%%%%%%%%%%%%%%%%%%%%%%%%%%%%%%%%%%%%%%%%%%%%%%%%%%%%%%%%

\end{document}